# Wavelength-multiplexed massively parallel diffractive optical information storage and image projection

Che-Yung Shen[1,2,3†], Yuhang Li[1,2,3†], Cagatay Isil[1,2,3†], Jingxi Li[1,2,3], Leon Lenk[4], Tianyi Gan[1,3], Guangdong Ma[1,2,3], Fazil Onuralp Ardic[1,2,3], Mona Jarrahi[1,3] and Aydogan Ozcan[1,2,3*]

[1]Electrical and Computer Engineering Department, University of California, Los Angeles, CA, 90095, USA

[2]Bioengineering Department, University of California, Los Angeles, CA, 90095, USA

[3]California NanoSystems Institute (CNSI), University of California, Los Angeles, CA, 90095, USA

[4]Department of Computer Science, University of California, Los Angeles, CA, 90095, USA

[†]These authors contributed equally to the work

[*]Correspondence to: ozcan@ucla.edu

## Abstract

We introduce a wavelength-multiplexed massively parallel diffractive information storage platform composed of dielectric surfaces that are structurally optimized at the wavelength scale using deep learning to store and project thousands of distinct image patterns, each assigned to a unique wavelength. Through numerical simulations in the visible spectrum, we demonstrated that our wavelength-multiplexed diffractive system can store and project over 4,000 independent desired images/patterns within its output field-of-view, with high image quality and minimal crosstalk between spectral channels. Furthermore, in a proof-of-concept experiment, we demonstrated a two-layer diffractive design that stored six distinct patterns and projected them onto the same output field of view at six different wavelengths (500, 548, 596, 644, 692, and 740 nm). This diffractive architecture is scalable and can operate at various parts of the electromagnetic spectrum without the need for material dispersion engineering or redesigning its optimized diffractive layers. The demonstrated storage capacity, reconstruction image fidelity, and wavelength-encoded massively parallel read-out of our diffractive platform offer a compact and fast-access solution for large-scale optical information storage, image projection applications.

# 1 INTRODUCTION

Amid rapidly growing demands in data analysis and machine-learning applications, the surge in data generation and storage necessitates innovative solutions that enhance both storage density and operational efficiency. While conventional digital storage systems based on magnetic media, such as hard disk drives, are widely employed, challenges have arisen around increasing costs, limited lifespan, and relatively slow access times. These issues are especially critical for long-term data storage, prompting researchers to investigate more efficient and durable storage methods. Optical holographic storage systems have received attention as viable options due to their potential for high data density and quick access times(*1–6*).

In holographic data storage, information is encoded through a coherent beam using optical interferometry or a spatial light modulator (SLM). Recent advances in metasurface technologies have also revived interest in optical data storage solutions(*7–11*). These subwavelength metallic and/or dielectric structures exploit the spatial and polarization multiplexing capabilities of metasurfaces(*12*), enabling the development of compact and efficient storage systems. Additionally, various nano-fabrication techniques such as two-photon polymerization(*13, 14*), thermal scanning-probe lithography(*15, 16*) and nanoscale volume holograms(*17, 18*) have further advanced high-capacity, compact storage designs with feature sizes down to nanometers. These technological breakthroughs signal a paradigm shift in high-capacity optical displays, information encryption, and data storage. Despite these advancements, optical storage technologies continue to face challenges, including large area fabrication for storage capacity(*9*), limited bandwidth and channel crosstalk that introduces noise to the restored images/data(*19, 20*).

Here, we introduce a wavelength-multiplexed, massively parallel diffractive optical storage system, utilizing a series of spatially-optimized dielectric diffractive layers, cascaded to each other to collectively form a compact diffractive optical medium, as illustrated in **Fig. 1**. Engineered through deep learning-based structural optimization(*21–25*), the diffractive optical storage design functions to store information and project the corresponding images/data, each assigned to a unique wavelength, onto a fixed output field-of-view (FOV). The restored intensity images at the output FOV can be measured using an image sensor array, enabling read-out of the stored information through wavelength multiplexing. Using numerical simulations, we demonstrated that our optical storage system can restore over four thousand distinct image patterns and project them at the output FOV at different wavelengths with high image quality and minimal crosstalk, with an average peak signal-to-noise ratio (PSNR) of >48 dB for the output images. This diffractive storage design also has a compact footprint that axially spans ~$1260\lambda_m$, where $\lambda_m$ represents the average wavelength of the illumination spectral band. Furthermore, we experimentally demonstrated a two-layer diffractive design that stored six distinct patterns and projected them onto the same output FOV at six different wavelengths, i.e., 500 nm, 548 nm, 596 nm, 644 nm, 692 nm, and 740 nm.

Through both numerical simulations and experiments, we demonstrated that our wavelength-multiplexed diffractive optical storage design could operate across different parts of the electromagnetic spectrum without the need for re-optimizing the diffractive layers or material dispersion engineering. These diffractive designs can also be combined with other multiplexing approaches, such as illumination angle(*2*, *11*, *26*), shift(*27*) and polarization multiplexing(*10*, *28*), to further boost their storage capacity. The presented approach offers new opportunities for designing advanced optical information storage platforms, potentially covering various applications such as image projection and holographic displays.

## 2 RESULTS

**Design of wavelength-multiplexed diffractive optical storage and image projection**

**Figure 1** depicts a schematic of our diffractive storage and image projection framework, illuminated by a spatially coherent light incorporating multiple wavelengths, i.e., $\{\lambda_1, \lambda_2, \ldots, \lambda_{N_w}\}$, arranged from the longest to the shortest wavelength, to optically synthesize/reconstruct the stored images corresponding to each wavelength in the illumination; $N_w$ refers to the number of distinct illumination wavelengths, each assigned to a unique target image/pattern. Following the input illumination aperture, the diffractive storage system consists of $K$ successive modulation layers made of dielectric materials. Each diffractive layer is structured with the same number of diffractive features that have a lateral size of $\sim \lambda_{N_w}/2$ and a trainable/learnable thickness, providing a phase modulation range covering 0-2π for all wavelengths in the input illumination. These layers, including the input/output planes, are connected through light diffraction in air, which can also be replaced by a dielectric medium (e.g., a polymer or glass film). The uniform input optical fields $\{i_w\}$ ($w \in \{1,2,\ldots,N_w\}$) are processed by the diffractive storage model to produce output complex fields $\{o_w\}$, one for each wavelength $\lambda_w$. The resulting output intensity profile at each illumination wavelength is captured by a monochrome image sensor, yielding an intensity measurement $O_w$. Each wavelength is designed to display a unique (randomly selected) pattern from the CIFAR100 dataset, enabling the storage and projection of $N_w$ distinct images across the same diffractive optical model (see the Materials and Methods section for further details). These diffractive models are optimized using error-backpropagation and stochastic gradient descent(*29–32*, *28*, *33*) methods, aiming to minimize a custom loss function $\mathcal{L}$, based on the mean-squared error (MSE) between the projected intensity images and their corresponding ground truths across all the illumination wavelengths (refer to the Materials and Methods section for further details).

**Figure 2** illustrates the performance of this diffractive design framework for storing $N_w = 4096$ images (25×25 pixels), each of which corresponds to a distinct visible wavelength $\{\lambda_1, \lambda_2, \ldots, \lambda_{N_w}\}$, uniformly distributed within a range of $\lambda_1 = 750$ nm to $\lambda_{4096} = 400$ nm. This design is composed of $K = 8$ diffractive layers, each including 800×800 diffractive features, with a lateral pitch of $0.5\lambda_{N_w}$ per feature. The optimized thickness profiles of the resulting diffractive layers are presented in **Fig. 2A.** The reconstruction quality of the stored images at the output FOV across different wavelength channels $\{\lambda_1, \lambda_2, \ldots, \lambda_{N_w}\}$ was evaluated using the PSNR metric, as shown in the left part of **Fig. 2B.** The system achieved a mean PSNR of 45.29 ± 1.52 dB (standard deviation), calculated using $N_w = 4096$ stored images and their ground truth counterparts, demonstrating the accurate storage/reconstruction performance of the presented diffractive optical approach.

The accuracy of this diffractive storage model was further improved through a dynamic channel-swapping technique that we devised. First, we arranged the stored images of this initial design in descending order based on their output PSNR values. These images were then sequentially reassigned to the wavelength set $\{\lambda_1, \lambda_2, ..., \lambda_{N_w}\}$, starting from the longest wavelength. This image-wavelength reassignment strategy ensures that images with more texture and higher resolution features (i.e., statistically more difficult to project images) were assigned to shorter illumination wavelengths, which offer a higher degree of freedom for the diffractive model to effectively restore the data. Subsequently, the trained diffractive design was further fine-tuned using these new image-wavelength pairings/assignments. Further details of this strategy can be found in the Materials and Methods section. After this dynamic channel-swapping technique, the average PSNR statistically significantly improved to 48.01 ± 0.57 dB. Furthermore, the standard deviation of the output PSNR values across all image-wavelength pairs was reduced by ~2.67-fold, demonstrating the success of our dynamic channel swapping strategy for providing enhanced image reconstruction/read-out accuracy and statistically improved consistency across all the stored images, as illustrated in the right panel of **Fig. 2B**. In addition to these quantitative results, some examples of the stored and reconstructed images are visualized in **Fig. 2C**, further validating the accuracy and uniformity of the stored images across different wavelength channels.

**Impact of the number (*N*) of diffractive features on wavelength-multiplexing scalability in diffractive optical storage models**

The information processing and storage capacity of the presented diffractive approach depends on the total number of diffractive features ($N$) available, the number of image-wavelength pairs ($N_w$) and the total number of pixels at the output FOV ($N_o$)(*34, 35*). To empirically explore the relationships between these parameters, we trained various diffractive storage models ($K = 8$) with $N = \{0.5N_oN_w, N_oN_w, 2N_oN_w\}$ for different $N_w$ values $\{128, 288, 512, 800, 1568, 2048, 4096\}$. The average output PSNR values resulting from these diffractive storage models across different $N$ and $N_w$ configurations are shown in **Fig. 3A**. This comparative analysis highlights that improving the degrees of freedom of the diffractive models, by increasing $N$, substantially enhances the overall output image quality of the diffractive model; for example, the average output PSNR for each case of $N_w$ from 128 to 4096 is reported as <30 dB when using $N = 0.5N_oN_w$, which increased to >48 dB when using $N = 2N_oN_w$. Notably, the average PSNR drops from 75.91 to 48.01 dB when $N_w$ increases from 128 to 4096, which can be attributed to the relatively increased spectral crosstalk resulting from reduced wavelength separation between the consecutive channels. **Figure 3 (B and C)** showcases exemplary output images from the diffractive designs with $N = 2N_oN_w$ and $N = 0.5N_oN_w$ for $N_w$ = 128 or 4096 image-wavelength pairs. This visual comparison also shows that the reconstructed images of the diffractive model with $N = 2N_oN_w$ exhibit higher contrast and less speckle noise compared to those with $N = 0.5N_oN_w$. These results affirm that, by scaling the number of trainable diffractive features in proportion to the number of wavelength channels, we can effectively maintain the quality of the projected output images and improve the wavelength multiplexing capability of the diffractive storage system.

**Impact of material dispersion on the image projection performance of wavelength-multiplexed diffractive optical storage models**

A notable characteristic of our wavelength-multiplexed diffractive optical storage system is that its operation does not rely on material dispersion engineering of the diffractive layers. To evaluate the impact of material dispersion on the performance of the presented wavelength-multiplexed diffractive system, we compared the original $N_w$ = 128 diffractive storage designs using N-BK7 glass dispersion in **Fig. 3B** to dispersion-free diffractive models exhibiting a constant refractive index (~1.5) across all the illumination wavelengths, which serve as "Gedanken" comparative experiments. These new dispersion-free diffractive designs with $N_w$ = 128 were optimized using the same physical configurations and training hyperparameters as those employed for the diffractive models reported in **Fig. 3B**. **Figure 4** summarizes the average PSNR and the corresponding standard deviation values for two representative configurations, $N = \{0.5N_oN_w, 2 N_oN_w\}$. Across these comparisons, the dispersion-free designs achieve a reconstruction performance that is statistically comparable to that of the N-BK7-based implementations. Both sets of designs exhibit the same performance trend as the number of diffractive features ($N$) increases, reaching peak PSNR values of >75 dB for the diffractive storage designs using $N=2N_oN_w$. Furthermore, as shown in **Fig. 4**, statistical analysis of the performance distributions yields p-values of >0.85 when comparing the diffractive storage designs using N-BK7 glass dispersion with those using a dispersion-free "Gedanken" material, indicating no statistically significant difference between the diffractive designs with or without material dispersion.

These results indicate that reconstruction fidelity is predominantly determined by the learned diffractive phase profiles rather than by the material's explicit wavelength-dependent refractive index. During the optimization of each diffractive design, the training loss function jointly evaluates reconstruction errors across all wavelength channels, guiding the diffractive layers toward a solution that balances performance uniformly across the entire spectral set. The observations in **Fig. 4** confirm that wavelength-multiplexed diffractive storage can be realized using a broad class of optical materials with diverse dispersion properties, including those with nearly flat dispersion curves. Collectively, these analyses demonstrate that our diffractive storage architecture is adaptable to various phase modulation elements and substrates, and can operate effectively across different regions of the electromagnetic spectrum.

**Output power efficiency of wavelength-multiplexed diffractive optical storage models**

Presented diffractive data storage designs were optimized without any constraints on the output power efficiency, exhibiting relatively low diffraction efficiencies, generally under 0.01%. To design a diffractive optical model with higher output efficiencies, we introduced an additional loss term into our training loss function(*31, 22, 36*) to balance the tradeoff between the accuracy and power efficiency of the system (see the Materials and Methods section for details). In **Fig. 5**, we report a quantitative analysis to explore the tradeoff between the multiplexed output image projection performance and the output diffraction efficiency of diffractive optical storage models. For this comparison, we designed various diffractive models with $N = \{0.5N_oN_w, N_oN_w, 2N_oN_w\}$ for $N_w$=128 image-wavelength pairs using different diffraction efficiency thresholds during the training, which were incorporated to set the final power efficiency for each design (see the Materials and Methods section for details). **Figure 5A** quantifies the tradeoff between the output image quality and the output diffraction efficiency for the models with different numbers of optimizable features, i.e., $N = \{0.5N_oN_w, N_oN_w, 2N_oN_w\}$. In the case of $N = 0.5N_oN_w$, when the diffraction efficiency penalty is factored in and the efficiency is increased from 0.008% to 0.1%, the image quality declines, with the average PSNR dropping from 75.91 dB to 21.23 dB. In comparison, when $N$ is doubled to $N_oN_w$, the diffractive storage systems can be designed with output efficiencies of up to ~4%, while still projecting output images with an average PSNR above 20 dB. Furthermore, if the number of optimizable features in the diffractive model is increased to $2N_oN_w$, the design can maintain a similar output image reconstruction performance with diffraction efficiencies as high as ~10%. Examples of the image projections for some of these diffractive models with varying diffraction efficiencies are shown in **Fig. 5B**. These examples reveal that as the diffraction efficiency increases, there is only a minor reduction in image fidelity for the $N = 2N_oN_w$ design. **Figure 6** also illustrates a similar analysis incorporating different diffractive designs with $N_w = \{128, 288, 512\}$ and a fixed $N/N_w$ ratio. These analyses reported in **Figs**. **5** and **6** help us better understand the image quality-power efficiency tradeoff for diffractive storage designs and how this tradeoff can be tailored by various design parameters, including $N$ and $N_w$.

**Impact of the phase bit depth ($B$) on the accuracy of wavelength-multiplexed diffractive storage models**

Diffractive storage models are affected by fabrication restrictions, including the limited axial resolution for the thickness features on each layer, which impacts the phase bit depth that is available. To investigate the impact of limited phase bit depth on the performance of a diffractive optical storage model, we limited the phase bit depth, $B$, at each diffractive layer to be $\{16, 8, 6, 4\}$. As illustrated in **Fig. 7A**, the image storage and projection performance of a diffractive model trained using 16-bit phase quantization for each diffractive feature (i.e., $B_{train} = 16$) gradually decreases with the reduction in phase quantization levels during testing, $B_{test} = \{16, 8, 6, 4\}$. For example, for the diffractive design trained/optimized with $B_{train} = 16$, the average output PSNR was 75.91 dB for $B_{test} = 16$, and it decreased to 17.83 dB for $B_{test} = 4$. However, when a diffractive optical storage model was trained using $B_{train} = 4$ and tested with the same bit depth (i.e., $B_{test} = B_{train} = 4$), the average output PSNR reached 25.12 dB. This analysis, along with the visual examples presented in **Fig. 7B**, quantifies the impact of lower phase quantization levels on the image reconstruction accuracy of the diffractive storage framework and demonstrates how incorporating such phase bit depth limitations into the training phase can enhance performance, as shown in **Fig. 7** for $B_{test} = B_{train} = 4$.

**Experimental validation of wavelength-multiplexed diffractive optical storage**

We experimentally demonstrated our wavelength-multiplexed diffractive storage and image projection system within the visible spectrum. Using the same device architecture and training procedures as the numerical model shown in **Fig. 2**, we designed a wavelength-multiplexed diffractive storage system capable of storing six desired image patterns and projected them onto an output FOV at six distinct wavelengths: 500 nm, 548 nm, 596 nm, 644 nm, 692 nm, and 740 nm. For the test patterns, six images from the MNIST dataset were chosen and resized to 15 × 15 pixels. Additionally, we employed an *in-situ* learning strategy(*37*) in our experiments to minimize the impact of misalignments and other physical imperfections that commonly occur when transferring digitally optimized optical designs to a physical setup (see the Materials and Methods section). The experimental setup, shown in **Fig. 8A**, comprised a multispectral illumination source, two diffractive layers ($L_1$ and $L_2$), implemented using phase-only SLMs, and a CMOS image sensor to capture the output intensity distributions (see the Materials and Methods section for details). A photograph of the setup is provided in **Fig. 8B**, and the designed diffractive phase profiles, projected on the SLMs, are shown in **Fig. 8C**.

**Figure 8D** presents our experimental results achieved with in-situ optimization at six illumination wavelength channels alongside their simulated/numerical counterparts and the corresponding target patterns. During these experiments, unknown/uncontrolled optical misalignments and other hardware imperfections(*38*) degraded the initial output images as shown in the results *before* the in-situ learning was applied; see **Fig. 8F**. To improve the output image reconstruction performance and mitigate the impact of such misalignments/imperfections in the optical hardware, we implemented in-situ training directly on the optical hardware(*37, 39–41*). For this in-situ optimization phase, we adopted a sequential approach, optimizing the diffractive layers for each illumination wavelength, one by one. After training on one wavelength, the resulting optimized parameters were used to initialize the training for the next illumination wavelength. As evaluated by the Pearson Correlation Coefficient (PCC) values reported in **Fig. 8E**, our in-situ training approach led to significant performance improvements for all the illumination wavelengths during the in-situ training phase. Crucially, the visual differences between the initial results before the in-situ training and the final image reconstruction results after the in-situ training of the diffractive model hardware, as visualized in **Fig. 8 (D and F)**, show major improvements: the pre-training outputs exhibit pronounced inter-channel crosstalk due to hardware/setup imperfections and misalignments, whereas the post-training results show markedly reduced crosstalk and enhanced image quality. After completing the sequential in-situ training across all six illumination wavelengths, we experimentally tested the optimized/final system hardware using the in-situ-trained diffractive layers that were fixed for all the illumination wavelengths. The experimental measurements, shown in **Fig. 8D**, corresponding to this in-situ optimized and fixed optical hardware align well with our numerical simulations, which confirm the capability of the proposed diffractive optical system to store multiple images in different wavelength channels and reconstruct them accurately using illumination wavelength encoding.

**3 DISCUSSION**

One of the attractive features of optical systems is their inherent parallelism, allowing for multiplexed writing and reading of data, which offers the potential for high data throughput. In our wavelength-multiplexed diffractive models, the stored image data can be accessed by, for example, using a broadband light source with a monochrome image sensor equipped with a tunable spectral filter, or by employing a wavelength-scanning light source with a monochrome image sensor to read out the multiplexed images. We want to emphasize that our diffractive storage is *not* limited to sequential read-out. This framework can be adapted for snapshot measurements by incorporating the functionality of spectral filter arrays directly onto the diffractive processor, as demonstrated earlier(*42, 43*). This adaptation allows the projected patterns at the output of the diffractive processor to be segmented according to a virtual filter-array pattern (without the need for an actual filter-array fabrication), enabling a monochrome sensor to capture signals, in a *snapshot*, from multiple stored patterns. After applying a standard image demosaicing process, all the encoded patterns corresponding to their illumination wavelength channels can be retrieved from a single intensity image. This snapshot and simultaneous data read-out approach would need a larger pixel-count monochrome image sensor to match the required pixel count at the output plane, i.e., $N_o N_w$.

The presented wavelength-multiplexed diffractive optical information storage approach was shown to reconstruct over four thousand distinct image patterns, correctly projecting the stored patterns at the output FOV with high image quality and minimal crosstalk (with an average PSNR of >48 dB). Considering the thin optical volume of the diffractive designs, each analog image inference step takes < 3 ps from the illumination aperture to the output detector plane. For wavelength-multiplexing capability, a key practical consideration is how densely the illumination wavelength channels can be arranged in the spectral domain while maintaining high-fidelity output image reconstruction. To address this question, we further analyze the image reconstruction performance as a function of wavelength channel spacing, along with different phase quantization levels for the diffractive features. Here, the wavelength channel spacing $\Delta\lambda$ is defined as the total wavelength range that is available ($\beta_w$) divided by the number of multiplexed wavelength channels ($N_w$), i.e., $\Delta\lambda = \beta_w/N_w$. In this analysis, the total wavelength range ($\beta_w$) was set as 2000 pm, while $\Delta\lambda$ was selected from [1000, 100, 10, 1] pm, and the phase bit depth of the diffractive features ($B_{test} = B_{train}$) was tested over the range [10, 8, 6]. As shown in **Fig. S1** in the **Supplementary Material**, the resulting analyses indicate a clear trade-off between the wavelength channel spacing and the reconstruction fidelity. With $\Delta\lambda \geq 100$ pm, the reconstructed image quality remains high across all phase quantization levels, indicating minimal inter-channel crosstalk and good output image fidelity. When $\Delta\lambda = 1$ pm, a larger degradation in performance is observed, which can be attributed to the finite spectral selectivity of the diffractive storage design. Notably, for the diffractive designs with $\Delta\lambda=10$ pm, which is comparable to the wavelength precision and linewidth of state-of-the-art narrow-linewidth laser sources, the diffractive designs continue to exhibit successful image reconstruction, albeit with reduced PSNR values and increased error variance. These results highlight a limit imposed by the illumination wavelength precision and the diffractive features' phase quantization level, providing a design guide/map for selecting an appropriate wavelength channel spacing in practical wavelength-multiplexed diffractive optical storage systems.

To further improve this diffractive information storage framework, additional multiplexing strategies(*11*), such as illumination angle(*2, 26*), polarization(*10, 28*) and shift multiplexing(*27*), can be employed to expand the storage capacity. By integrating combinations of these multiplexing techniques, it is possible to reduce crosstalk further and significantly increase the number of retrievable images within the system.

Another practical consideration for the presented wavelength-multiplexing framework is the robustness of image reconstructions against uncontrolled variations in the probe wavelength. To quantify this effect, we evaluated the reconstructed image quality under shifted wavelengths that deviated from the corresponding desired wavelength channel. For each wavelength channel, the average PSNR is computed over 20 probe wavelengths uniformly sampled within a wavelength shift range [-0.1Δλ, 0.1Δλ], and the standard deviation values across these wavelength shifts are analyzed. As shown in **Fig. S2A** in the **Supplementary Material**, the image reconstruction quality is maximized when the probe wavelength matches the design wavelength, as expected, and it gradually degrades as the illumination wavelength detunes. Specifically, the PSNR value decreases from ~48.01 dB at the nominal wavelength to ~38.7 dB across the tested wavelength shift range. A representative middle channel performance shown in **Fig. S2B** in the **Supplementary Material** further illustrates this behavior, exhibiting a well-defined PSNR peak at the design wavelength, followed by a smooth and monotonic decline with increasing wavelength detuning. Importantly, successful image reconstruction is maintained within a finite wavelength deviation window, beyond which inter-channel phase mismatch and reduced spectral selectivity of the diffractive layers lead to noticeable image degradations. As shown in **Fig. S2B** in the **Supplementary Material**, this acceptable spectral window is on the order of a few tens of picometers, which is comparable to the wavelength stability and linewidth of state-of-the-art narrow-linewidth laser sources. These results suggest that while precise wavelength control is beneficial for optimal performance, the presented diffractive architecture is not strictly limited to an idealized single-wavelength probe and can accommodate realistic wavelength drifts without a major loss of reconstruction fidelity.

We also analyzed the resilience of the presented wavelength multiplexing framework to potential misalignments between the diffractive layers and the input/illumination plane; see **Figs. S3** and **S4** in the **Supplementary Material**. For this analysis, we employed an *in silico* vaccination strategy, introducing random lateral ($D_x \sim U(-\Delta_{xy,tr}, \Delta_{xy,tr})$, $D_y \sim U(-\Delta_{xy,tr}, \Delta_{xy,tr})$) and axial ($D_z \sim U(-\Delta_z, \Delta_z)$) displacements to the diffractive layers during the training phase. The baseline model, trained without any vaccination ($\Delta_{xy,tr} = \Delta_z = 0$), exhibits a rapid decline in the output image PSNR as the testing misalignment increases, resulting in severe image degradations. However, the diffractive models trained with vaccinated lateral shifts ($\Delta_{xy,tr}$=20 nm and 40 nm) demonstrate remarkable robustness, maintaining high-fidelity image reconstruction even under lateral displacements up to ±40 nm. As shown in **Fig. S3A** in the **Supplementary Material**, when $\Delta_{xy,test}$=80 nm, the PSNR value for the vaccinated design ($\Delta_{xy,tr}$=40 nm) remained at ~29.2 dB, while the PSNR value for the unvaccinated baseline design drops to ~21.9 dB. A similar trend is also observed for axial misalignments, as shown in **Fig. S4** in the **Supplementary Material**, where the diffractive models vaccinated with axial uncertainty ($\Delta_{z,tr}$=100 nm and 200 nm) effectively mitigate the performance degradation associated with inter-layer spacing errors or misalignments. Crucially, this enhanced robustness incurs only a negligible penalty on the peak image quality under ideal alignment conditions. These findings reveal that incorporating stochastic positional noise during the training process effectively desensitizes the diffractive storage performance to mechanical variations or imperfections, thereby partially relaxing the strict fabrication tolerances of these wavelength-multiplexed diffractive storage systems.

Inherent data security is also one of the crucial aspects of diffractive information storage systems. In our design, accessing the stored information within a diffractive model requires the precise knowledge of both the specific illumination wavelengths and the input aperture, which effectively function as "keys" to our diffractive information storage system. To further strengthen the security of the storage platform, an additional diffractive layer can be introduced as another optical key(*44, 45*). This diffractive layer could add an extra level of encryption, ensuring that data retrieval is possible only when this unique component (i.e., the diffractive key) is correctly implemented. Consequently, duplicating or replicating the stored data would demand not only advanced fabrication capabilities to create the physical diffractive key but also detailed access to the specific optical read-out setup. Therefore, this multi-layer approach could significantly enhance data security and privacy, making unauthorized access/read-out highly challenging.

In conclusion, we developed a diffractive optical information storage system capable of storing thousands of distinct patterns and accurately projecting these patterns across a broad range of spectral bands, with the analog output data being accessible within a few picoseconds of light propagation through thin optical components. Our deep-learning-enabled approach effectively mitigates spectral crosstalk, ensuring high image quality across all the illumination wavelength channels. With appropriate nano-fabrication techniques, e.g., using two-photon polymerization-based 3D printing(*46–48*), projection lithography(*49, 50*), thermal scanning-probe lithography(*15, 16*), or electron beam lithography(*8, 51–53*), these multi-layer storage designs can be physically scaled to function at different parts of the electromagnetic spectrum without any material dispersion engineering or the need for redesigning the optimized diffractive layers. Furthermore, previous studies showed that laser-based glass writing platforms could provide a complementary fabrication pathway for implementing the presented diffractive optical storage systems within ultra-stable media(*6*). By leveraging femtosecond laser writing in quartz, such approaches demonstrated the feasibility of fabricating anisotropic, long-lived optical structures, which could further support the practicality and durability of the presented wavelength multiplexed diffractive storage framework.

Using SLMs as part of the same framework can also achieve re-writable optical information storage through re-programming of the diffractive layers for new data/images. We believe that these wavelength-multiplexed diffractive models hold significant potential for large-scale, rapid-access information storage, private communication and security-related applications.

**4 MATERIALS AND METHODS**

**Optical forward model of wavelength-multiplexed diffractive systems**

Our diffractive models consist of $K$ consecutive diffractive layers, each containing thousands of precisely positioned diffractive features at the wavelength scale. In the numerical forward model, these layers are treated as thin planar structures that modulate the incident coherent light with a complex transmission function. For any given $s^{\text{th}}$ diffractive feature on the $l^{\text{th}}$ layer located at $(x_s, y_s, z_l)$, its complex-valued transmission coefficient, dependent on the material thickness value $h_s^l$, can be expressed as:

$$t(x_s, y_s, z_l; \lambda) = \exp\left(\frac{-2\pi\kappa(\lambda)h_s^l}{\lambda}\right)\exp\left(\frac{-j2\pi(n(\lambda) - n_{\text{air}})h_s^l}{\lambda}\right) \quad (1).$$

Here, $n(\lambda)$ and $\kappa(\lambda)$ are the real and imaginary parts of the material's complex refractive index $\tilde{n}(\lambda)$, i.e., $\tilde{n}(\lambda) = n(\lambda) + j\kappa(\lambda)$. For all the numerical simulations of diffractive storage designs within the visible spectrum reported in this paper, we selected N-BK7 as the material of the diffractive layers(54). Since N-BK7 exhibits negligible absorption in the visible spectrum, $\kappa(\lambda)$ was assumed to be 0. The thickness value $h$ of each diffractive element is composed of two parts: a learnable thickness $h_{\text{learnable}}$ and a base thickness $h_{\text{base}}$, i.e.,

$$h = h_{\text{learnable}} + h_{\text{base}} \quad (2),$$

Here, $h_{\text{learnable}}$ is the tunable thickness value of each diffractive feature optimized during the training process and is constrained within the range $[0, h_{\max}]$; $h_{\text{base}}$ is a fixed value representing the base thickness that acts as the substrate support for the diffractive layer. For the diffractive storage designs within the visible spectrum, $h_{\max}$ was set as 664.5 nm, corresponding to a full phase modulation range (0 to $2\pi$) for the longest wavelength ($\lambda_1$). $h_{\text{base}}$ was empirically chosen as 700 nm.

For the experimentally validated diffractive storage and image projection design in **Fig. 8A**, as we utilized phase-only SLMs, we applied a similar model and treated the diffractive layer as a pure phase modulator, with a reflection coefficient of:

$$r(x_s, y_s, z_l; \lambda) = \exp(-j\phi(x_s, y_s; \lambda)) \quad (3).$$

Here, $\phi(x_s, y_s; \lambda)$ is the phase modulation value for the corresponding diffractive feature, optimized during the training process.

To numerically model free-space propagation of coherent light between the diffractive layers, we employed the Rayleigh-Sommerfeld scalar diffraction theory. In our numerical simulations, the diffraction process is formulated as a linear, shift-invariant operator with an impulse response. Each $s^{\text{th}}$ diffractive feature on the $l^{\text{th}}$ layer at $(x_s, y_s, z_l)$ is defined as the source of a secondary wave, generating a complex field at wavelength $\lambda$ given by the equation:

$$w_s^l(x, y, z; \lambda) = \frac{z - z_l}{(r_s^l)^2} \left( \frac{1}{2\pi r_s^l} + \frac{n}{j\lambda} \right) \exp\left( \frac{j2\pi n r_s^l}{\lambda} \right) \quad (4),$$

where $r_s^l = \sqrt{(x - x_s)^2 + (y - y_s)^2 + (z - z_l)^2}$. These secondary waves propagate to the next layer ($(l+1)^{\text{th}}$ layer), where they are spatially superimposed. Therefore, the optical field that reaches the $p^{\text{th}}$ diffractive feature in the $(l+1)^{\text{th}}$ layer, located at $(x_p, y_p, z_{l+1})$, can be computed by the convolution of the complex field from the previous layer with the impulse response function $w_s^l(x_p, y_p, z_{l+1}; \lambda)$. The resulting field is then modulated by the transmission function $t(x_p, y_p, z_{l+1}; \lambda)$ of the $(l+1)^{\text{th}}$ diffractive layer, which can be expressed as:

$$u_p^{l+1}(x, y, z; \lambda) = t(x_p, y_p, z_p; \lambda) \sum_s u_s^l w_s^l(x_p, y_p, z_{l+1}; \lambda) \quad (5).$$

In the numerical simulations of the diffractive storage designs, the spatial sampling period of the simulated complex fields was set to be half of the shortest wavelength, i.e., $0.5\lambda_{N_w} = 200$ nm, which also defines the lateral dimensions of a diffractive feature. As for the experimentally validated diffractive storage and image projection design in the visible spectrum, the spatial sampling period was set to be 4 μm, half of the SLM's pixel width.

For the diffractive storage design shown in **Fig. 2A**, which operates at $N_w$ = 4096 different illumination wavelengths, the sizes of the input aperture and the output FOV were both set to be ~69.56$\lambda_m$ × 69.56$\lambda_m$. The output FOV consists of $N_x \times N_y$ = 25 × 25 pixels, resulting in each output pixel having a size of ~2.78$\lambda_m$ × 2.78$\lambda_m$. We designed the diffractive model to possess 800 × 800 diffractive features per layer, resulting in a diffractive area of ~278$\lambda_m$ × 278$\lambda_m$ at each layer. Other diffractive storage designs analyzed in **Fig. 3A** used a similar architecture but varied in their number of diffractive features.

For the experimentally validated diffractive storage and image projection design shown in **Fig. 8A**, the output FOV was set as 2.4 mm × 2.4 mm, which is divided into 15 × 15 pixels, where each pixel has a size of 0.16 mm × 0.16 mm. The illumination wavelengths for the experimental validation were selected as [500, 548, 596, 644, 692, 740] nm. We designed the diffractive model to possess 800 × 800 diffractive features per layer, resulting in a diffractive area of 6.4 mm × 6.4 mm at each layer.

**Data preparation and other implementation details**

For the training of our numerically simulated diffractive storage models, we created target images by randomly selecting $N_w$ images from the CIFAR100 training dataset(*55*), each corresponding to a monochrome illumination at a specific wavelength chosen from $N_w$ predefined wavelengths.

To train and evaluate the experimentally validated diffractive storage and image projection design operating in the visible spectrum, we created target images by selecting six images from the MNIST training dataset, each corresponding to a monochrome illumination at a specific wavelength from six predefined wavelengths.

All numerical simulations and training procedures for our diffractive storage designs were implemented using Python (version 3.10.4) and JAX (version 0.4.1). We employed the Adam optimizer, utilizing the default parameters provided by the OPTAX library, with a learning rate set at 0.001 and a batch size of 128. The diffractive models were trained for 200 epochs on a workstation equipped with an Nvidia GeForce RTX 3090 GPU, an Intel Core i9-11900 CPU and 128 GB of RAM. The training of the 8-layer diffractive storage design with $N_w$ = 4096 shown in **Fig. 2A** took approximately 6 days to complete. It is important to emphasize that this training process of our diffractive storage design is a one-time effort, which is negligible compared to the practical lifetime of the stored information within the diffractive dielectric layers. The in-situ training was performed on a workstation with an Nvidia GeForce RTX 2080-Ti GPU, and an AMD Ryzen 5 1600 CPU; each in-situ learning epoch took on average 21 sec.

**Supplementary Materials:** This file contains:

- **Figs. S1-S4**.
- **Training loss function and performance analysis metrics**
- **Details of the experimental diffractive information storage & image projection system**

**Figures**

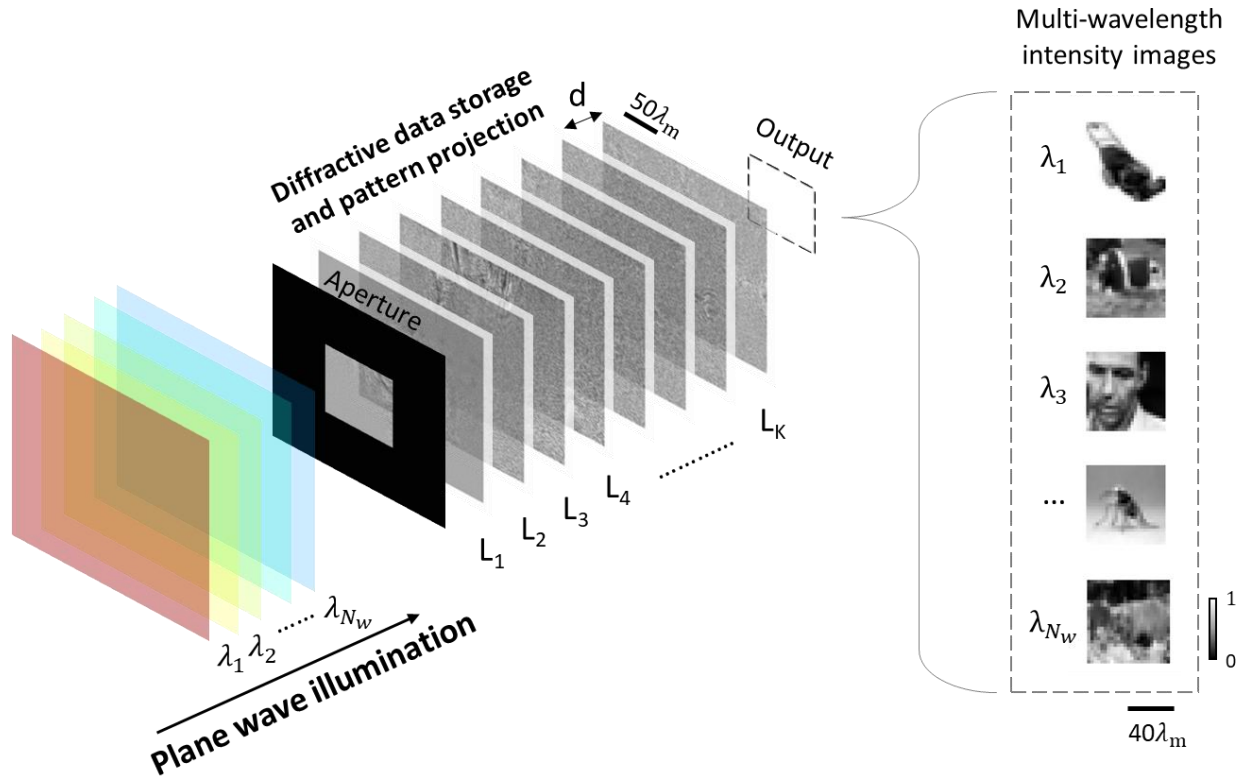

**Figure 1. Schematics of diffractive optical data storage and image projection.** Illustration of diffractive storage composed of $K$ diffractive layers, each comprising sub-wavelength phase elements jointly optimized through deep learning. The system is capable of storing and projecting $N_w$ distinct images/patterns when the corresponding wavelengths $\{\lambda_1, \lambda_2, \ldots, \lambda_{N_w}\}$ are incorporated in the illumination. We report the storage of over 4000 distinct patterns within a compact ($K \ll N_w$) diffractive design that axially spans $\sim 1260\lambda_m$, where $\lambda_m$ represents the average wavelength of the illumination spectral band.

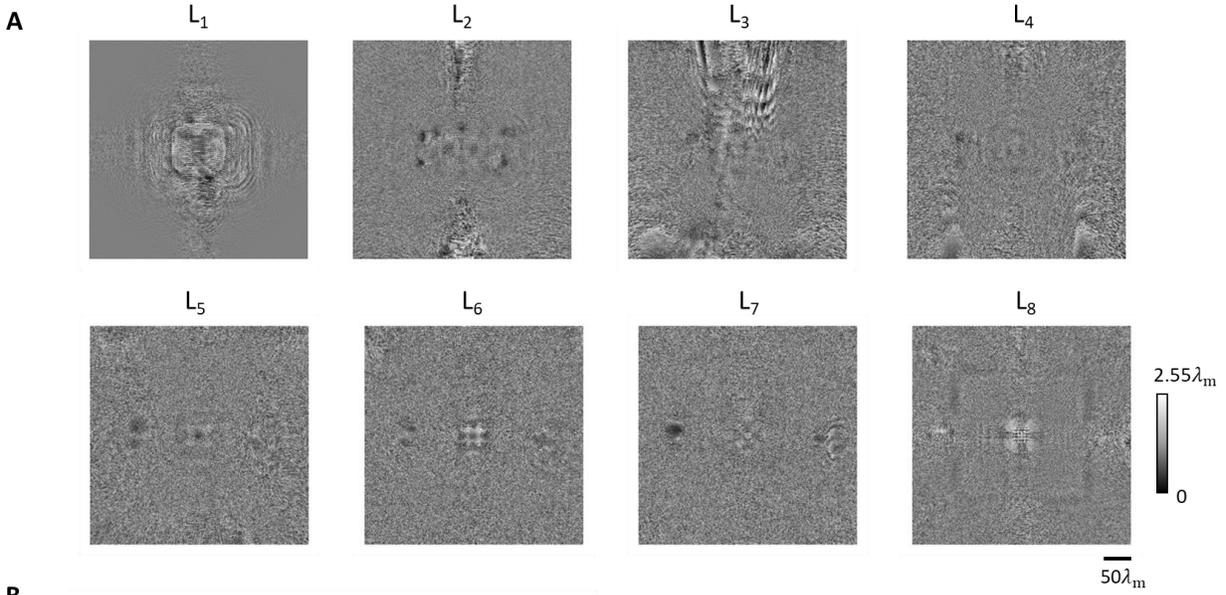

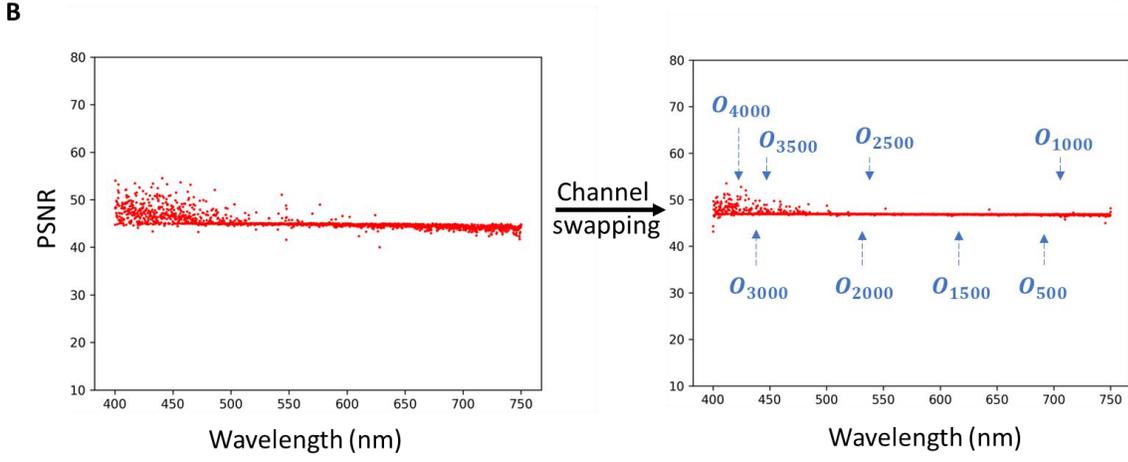

**Figure 2. Performance analysis of the diffractive storage design with $N_w = 4096$.** (A) Thickness profiles of the trained diffractive layers. (B) Output PSNR values of the diffractive outputs, each assigned to a distinct wavelength within a set of $N_w = 4096$ wavelength channels. The left part is the resulting performance based on training from scratch. The right part is the fine-tuning of this pretrained design using a dynamic channel swapping strategy, which improved the average output PSNR to 48.01 dB, while also reducing the standard deviation of the output PSNR values across all the image-wavelength pairs by ~2.67-fold compared to the design on the left. (C) Examples of the stored and reconstructed images ($O_w$) of the diffractive model, along with their ground truth counterparts and the corresponding performance evaluation with output PSNR values.

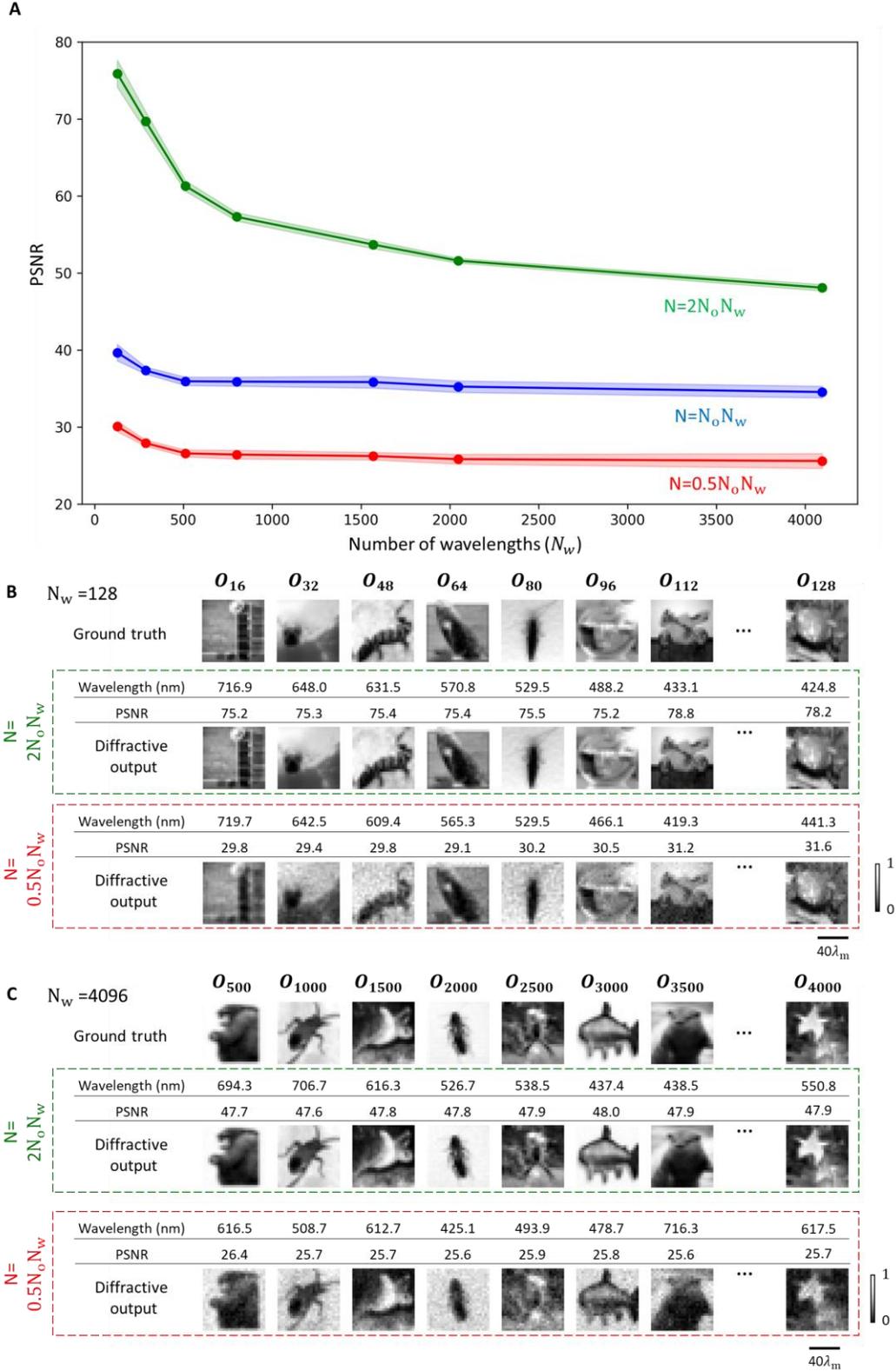

**Figure 3. Analysis of the impact of the number ($N_w$) of wavelength channels and the number ($N$) of diffractive features on the diffractive storage performance.** (A) Average output PSNR

values of different diffractive models using $N_w$=128 to 4096 and $N = \{0.5N_oN_w, N_oN_w, 2N_oN_w\}$. (B) Visualization of the diffractive outputs from the diffractive models designed under different numbers of trainable diffractive features ($N_w$=128). (C) Examples of the diffractive outputs, along with the ground truth images and the corresponding performance evaluation with the output PSNR values, from the diffractive models designed under different numbers of trainable diffractive features ($N_w$=4096).

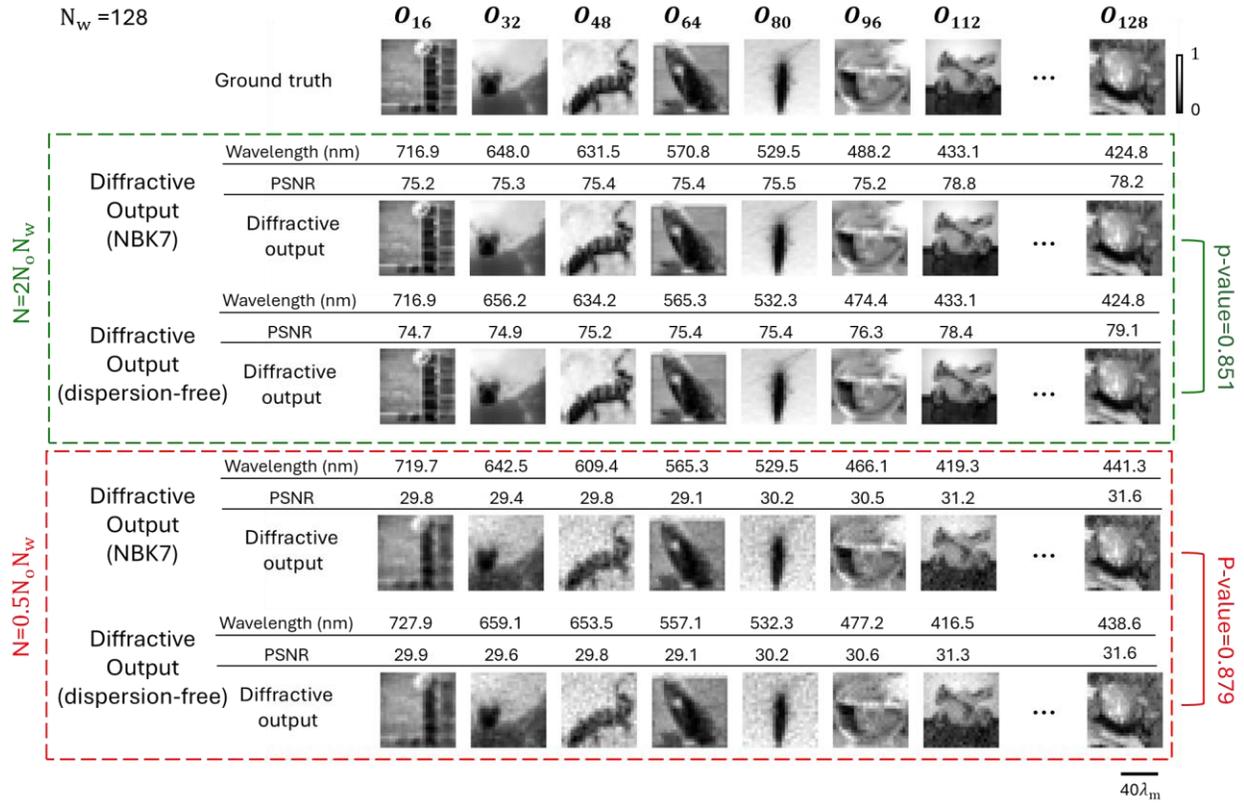

**Figure 4. The impact of material dispersion on the image projection performance of wavelength-multiplexed diffractive storage systems.** Exemplified images of diffractive storage designs ($N = 0.5N_oN_w, 2N_oN_w$) along with the calculated p-values between the diffractive storage models using N-BK7 glass dispersion and those using dispersion-free material designs ("Gedanken" experiments).

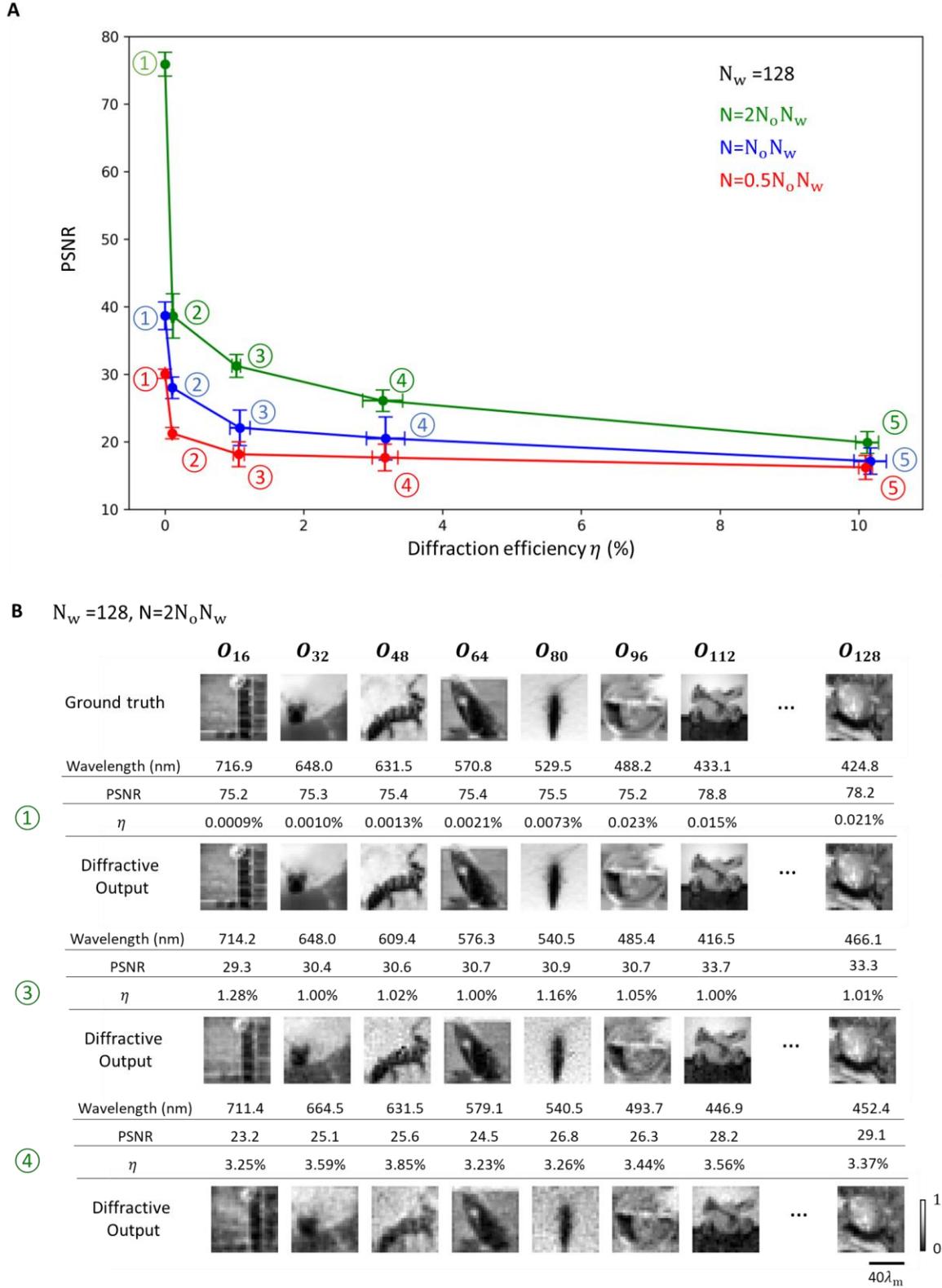

**Figure 5. Analysis of the tradeoff between the output diffraction efficiency ($\eta$) and the image quality of the diffractive models using the $N_w = 128$ design.** (A) PSNR values of the resulting

output patterns as a function of the diffraction efficiency. (B) Examples of the diffractive outputs, along with the ground truth images and the corresponding performance evaluation with the output PSNR and diffraction efficiency values from the $N_w$=128 diffractive storage designs trained under different efficiency thresholds.

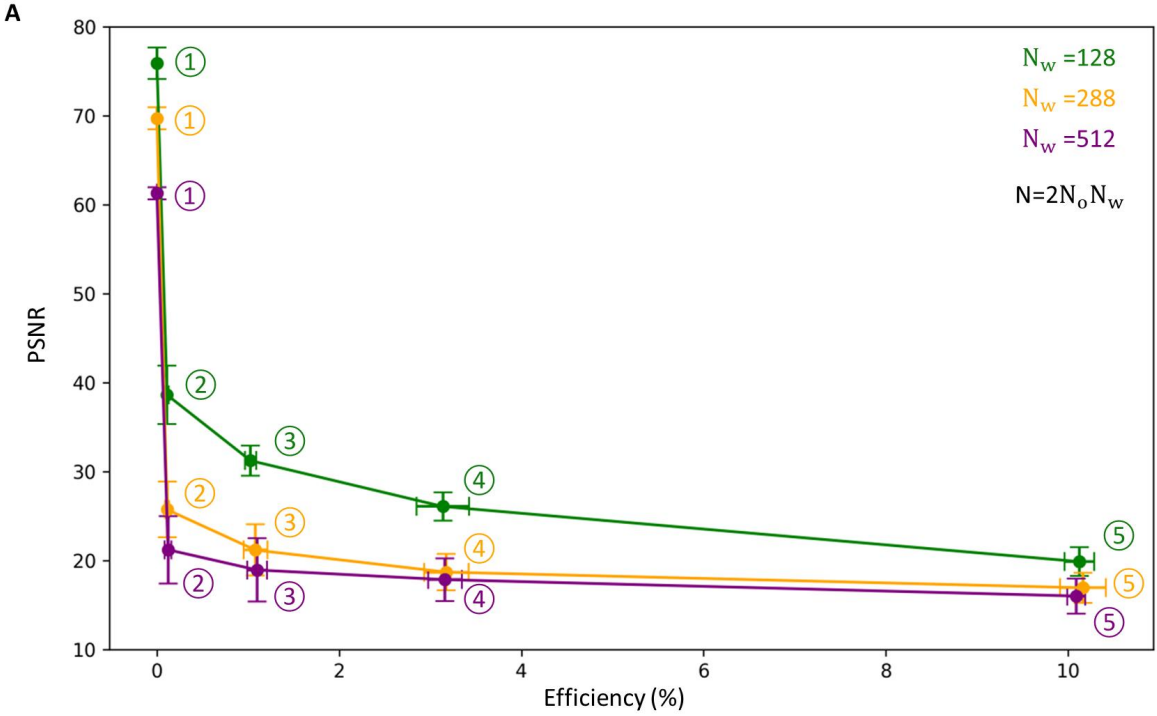

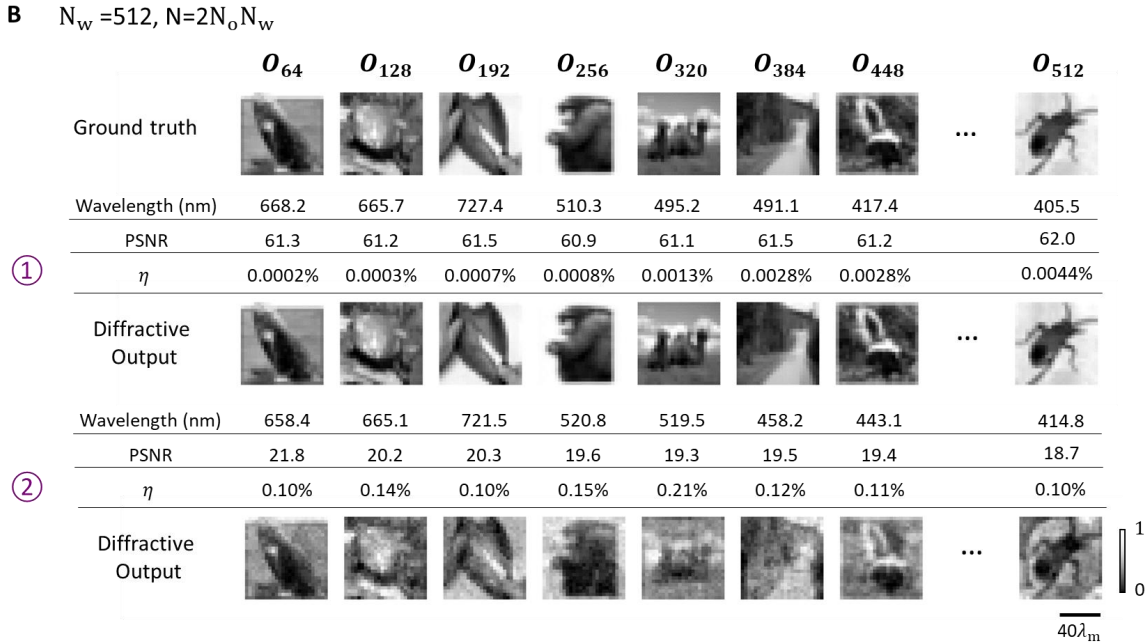

**Figure 6. Diffraction efficiency analysis of the diffractive storage designs under different numbers of wavelength channels.** (A) Output PSNR values of the resulting image projection models as a function of the output diffraction efficiency. (B) Examples of the diffractive outputs, along with the ground truth images and the corresponding performance with the output PSNR and diffraction efficiency values from the $N_w$=512 diffractive storage designs trained under different efficiency thresholds.

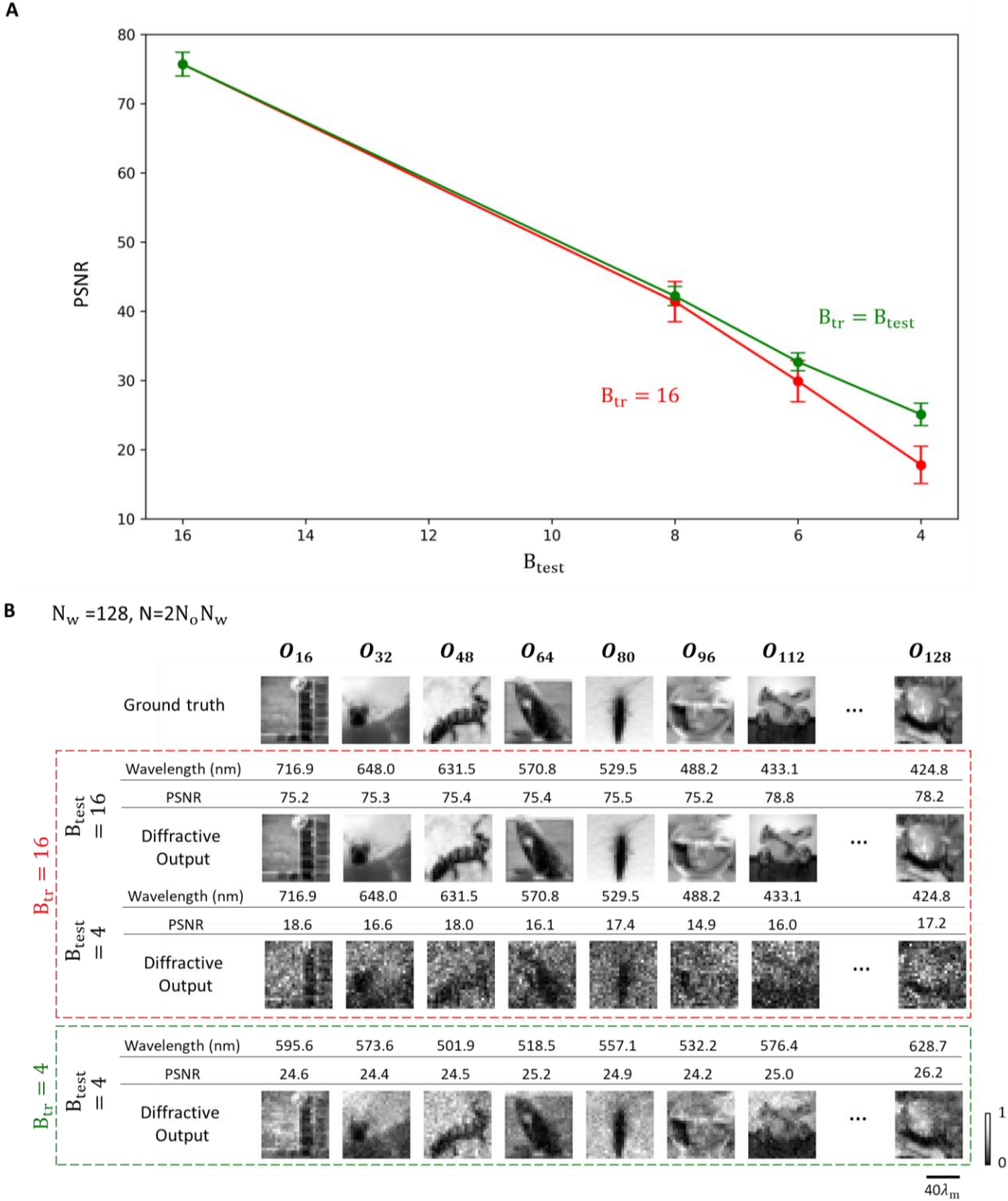

**Figure 7. Impact of phase quantization for the $N_w = 128$-channel diffractive storage design.** (A) Output PSNR values of the resulting image projection model as a function of the testing phase bit depth ($B_{test}$). (B) Examples of the diffractive outputs, along with the ground truth images and the corresponding performance evaluation with the output PSNR values from the $N_w$ =128 diffractive storage designs under different training ($B_{tr}$) and testing ($B_{test}$) phase bit depths.

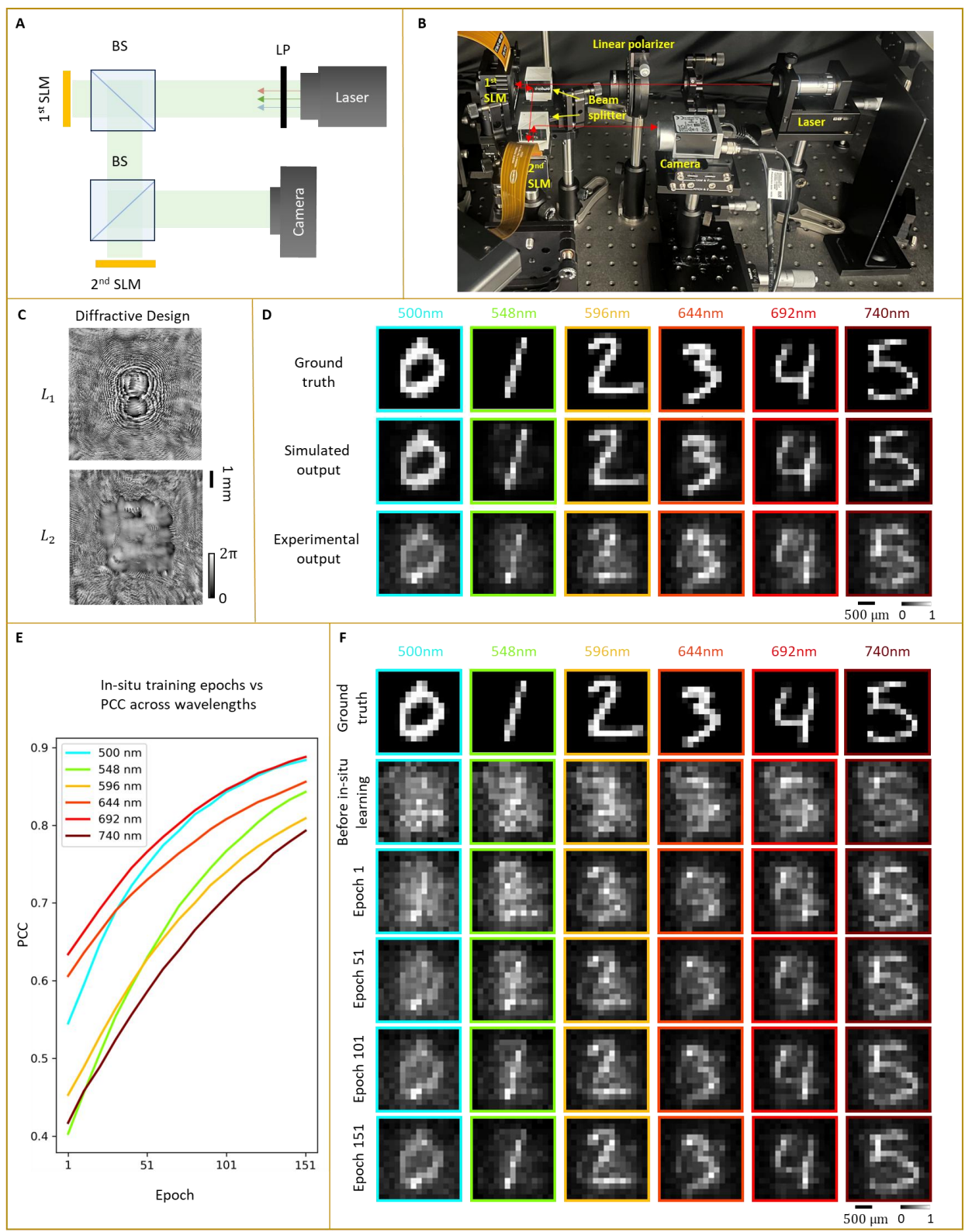

**Figure 8. Experimental demonstration of wavelength-multiplexed diffractive storage and image projection in the visible spectrum.** (A) Schematic of the diffractive image projection

system. (B) Photographs of the experimental setup. (C) Phase profiles of the optimized diffractive layers. (D) Numerically simulated and experimentally measured intensity patterns at the output plane, compared with the ground truth images. The color frames represent the corresponding illumination wavelengths, i.e., 500 nm, 548 nm, 596 nm, 644 nm, 692 nm, and 740 nm. (E) Evolution of the PCC values calculated between the diffractive outputs and the target images over 151 in-situ training epochs for six illumination wavelengths (500 nm, 548 nm, 596 nm, 644 nm, 692 nm, 740 nm). All the illumination wavelengths show monotonic PCC improvement as a function of the in-situ training epochs, as desired. (F) Representative diffractive output images at selected epochs. The top row shows the target digit patterns for each wavelength. Subsequent rows display the optical reconstructions before the in-situ learning as well as at epoch numbers 1, 51, 101, and 151 (during the in-situ learning). Progressive refinement of image quality is observed with increasing epochs, demonstrating the success of in-situ learning, which mitigates experimental imperfections and alignment issues.